\documentstyle[12pt,epsfig]{article}

\def\bg{\begin{eqnarray}}
\def\en{\end{eqnarray}}

\def\vr{\vec{r}}

\input{epsf.sty}
\def\framegraphics{\def\ifframe{\iftrue}}
\def\dontframegraphics{\def\ifframe{\iffalse}}
\def\drawgraphics{\def\ifdraw{\iftrue}}
\def\dontdrawgraphics{\def\ifdraw{\iffalse}}

\dontframegraphics
\drawgraphics

\newcommand{\be}{\begin{equation}}
\newcommand{\ee}{\end{equation}}
\newcommand{\bear}{\begin{eqnarray}}
\newcommand{\ear}{\end{eqnarray}}
\newcommand{\la}{\label}

\begin{document}


\vspace{2cm}
\centerline{\large Spin Dependent Parton Distributions in a Bound
Nucleon}
\vspace{.4cm}
\centerline{F.M. Steffens\footnote{Email
address:fsteffen@axpfep1.if.usp.br},
K. Tsushima\footnote{Email address:ktsushim@physics.adelaide.edu.au},
A. W. Thomas\footnote{Email address:athomas@physics.adelaide.edu.au}
and K. Saito\footnote{Email address:ksaito@nucl.phys.tohoku.ac.jp}}
\vspace{1cm}
\centerline{$^1$Instituto de F\'{\i}sica}
\centerline{Universidade de S\~ao Paulo}
\centerline{C.P. 66318}
\centerline{05389-970, S\~ao Paulo, Brazil}
\vspace{1cm}
\centerline{$^{2,3}$Department of Physics and Mathematical Physics}
\centerline{and}
\centerline{Special Research Center for the Subatomic Structure of
Matter}
\centerline{The University of Adelaide, SA 5005, Australia}
\vspace{1cm}
\centerline{$^{4}$Physics Division, Tohoku College of Pharmacy}
\centerline{Sendai, 981-8558 Japan}
\vspace{.3cm}

\begin{abstract}

The determination of the spin distribution functions for the neutron
requires an understanding of the changes induced by a nuclear medium. We
present the first estimates of the changes induced in the
spin-dependent,
valence parton distributions of the proton and neutron bound in $^3$He
and $^6$Li. For $^3$He the result is quite reassuring, in that the
off-shell
correction to $g_{1n}$ is very small. On the other hand, in $^6$Li the
corrections to the spin distributions in the bound proton are
sufficiently
large that they should be taken into account if one aims for a precision
of better than 10\%. We provide a simple parametrization of the
off-shell changes in the valence spin distributions of $^3$He and
$^6$Li, which can be used with any conventional estimate of nuclear
binding and Fermi motion corrections.
\end{abstract}

There is considerable interest in deriving the corrections to the
parton distributions of bound nucleons in order to finally understand
the nuclear EMC effect \cite{emcnuc,emcth}.
However, from the practical point of view
there is a more pressing need. In the light of the activity associated
with
the violation of the Ellis-Jaffe sum rule \cite{EJ},
also observed first by the
European Muon Collaboration (the ``EMC spin effect'') \cite{emcspin},
there has been a concentrated drive to determine $g_{1n}$ in order to
test
the Bjorken sum rule \cite{BJ}. This has led to a tremendous increase in
our knowledge of the proton and neutron spin structure,
both theoretically and experimentally \cite{smc}--\cite{D3}. On the
other hand,
all of the information about the neutron comes from nuclear targets --
so far
D, $^3$He and most recently $^6$Li.
It is therefore imperative to have control over the
corrections to the quark distributions for a bound nucleon.

{}For the deuteron the nuclear complications are sufficiently well
understood
that there have been several studies of the corrections to the neutron
spin distributions extracted from D data \cite{Deut}.
In this case it seems
that the off-shell corrections are under control, at least for $x \leq
0.7$.
However, when it comes to heavier nuclei, such as $^3$He, $^6$Li and
$^{15}$N,
the nuclear structure corrections may be of much more practical
importance.
Apart from this theoretical motivation, there is already some hint from
recent experiments that nuclear corrections to nucleon properties may be
significant in He. For example, recent
studies of the neutron charge form factor in $^3$He~\cite{nform}
suggest a possible difference between the measurement
in D and $^3$He \cite{exnform}.

In order to estimate the nuclear corrections to the parton
distributions in nuclei such as these,
at least at the present time when it is impossible to make
a complete QCD based calculation, we need two ingredients. First, we
need
a model for the nucleon from which we can calculate free nucleon parton
distributions that are in reasonable agreement with experimental data.
Second,
we need a theory based on the same quark model which can reproduce the
essential features of the structure of the nucleus of interest.

We consider first the issue of calculating the parton distributions
corresponding to a particular quark model. There
is a long history of studies of this kind, beginning with the work of
Le Yaouanc {\it et al.} \cite{strfn}, Jaffe \cite{RJ} and Parisi and
Petronzio \cite{PP} in the mid-70's.
The basic idea is to connect the twist-2 parton
distributions at a low scale with a valence dominated quark model -- an
idea exploited very successfully at the phenomenological level by
Gl\"uck {\it et al.} \cite{GRV}.
A major problem with early quark model calculations, namely poor
support, was solved about 10 years ago by formulating the
problem in such a way that energy-momentum conservation was guaranteed
{\it before} any approximation was made \cite{ST1}. Using that technique
it has been shown that, provided one allows for the hyperfine mass
splitting between the S=0 and S=1 spectator pairs, the MIT bag model can
give quite a good quantitative description of the observed valence
parton distributions \cite{tony91,me94a}.

The second essential ingredient has only been obtained quite recently.
The quark meson coupling model (QMC), which is based
on explicit quark degrees of freedom,
is ideally suited for this purpose \cite{gui}--\cite{Jen}.
In its simplest form it is also based on the MIT bag model, with the
interactions between different nucleons described by the exchange of
scalar and vector mesons in mean field approximation. It has
proven successful in reproducing the saturation properties of nuclear
matter as well as the binding energies and charge densities of finite,
closed shell nuclei. From the practical point of view, one
of the most attractive features of the model is
that it is not significantly more complicated than
Quantum Hadrodynamics (QHD)~\cite{qhd} -- even though the quark
substructure of hadrons is explicitly implemented.
A detailed description of the Lagrangian density, and the
mean-field equations of motion needed to describe a finite nucleus,
is given in Refs.~\cite{finite0,finite1,qmcapp}.

Let us briefly outline some key features of QMC needed in the present
calculation. For a bag centred at
position $\vr$ in a nucleus (with the coordinate origin taken at
the center of the nucleus), the Dirac equations for the quarks
in the nucleon bag are given by~\cite{finite0,finite1}
\be
\left[ i \gamma \cdot \partial_x - (m_q - V_\sigma(\vr)) - \gamma^0
\left( V_\omega(\vr) \pm \frac{1}{2} V_\rho(\vr) \right) \right]
\left(\begin{array}{c} \psi_u(x)\\ \psi_d(x)\\ \end{array}\right)
 = 0
\la{1}
\ee
The mean-field potentials at the centre of the bag are defined by
$V_\sigma(\vr) = g^q_\sigma \sigma(\vr),
V_\omega(\vr) = g^q_\omega \omega(\vr)$, and
$V_\rho(\vr) = g^q_\rho b(\vr)$, with $g^q_\sigma, g^q_\omega$
and $g^q_\rho$ being, respectively, the corresponding
quark and meson-field coupling constants. (Note that we have neglected
the variation of the scalar and vector mean-fields inside the nucleon
bag due to its finite size~\cite{finite0}.) For $^3$He,
which is too light for a simple shell model description,
we use the empirical density distributions~\cite{3he} and
calculate mean-field potentials in the nucleus using local density
approximation.
However, for $^6$Li we calculate the mean-field potentials
and proton and neutron density distributions in the QMC model
self-consistently, by solving Eqs.~(23) -- (30) of Ref.~\cite{finite1}.

The effect of the mean field potentials on the internal structure of the
nucleon can be totally absorbed into the
normalization constant, the quark eigenenergy, a small change in the bag
radius compared with free space
and the relative renormalization of the lower component of the
Dirac spinor:
\bear
\psi_q(\vec r', t) = N e^{-iE_q t/R}
\left(\matrix{j_0(xr'/R) \cr
              i\beta\vec\sigma\cdot\hat r j_1(xr'/R)}\right)
\frac{\chi_s}{\sqrt{4\pi}},
\la{2}
\ear
where $r'$ is the quark coordinate in the bag and
\bear
E_q &=& \Omega + R(V_\omega(\vr) \pm \frac{1}{2} V_\rho (\vr))
\quad{\rm for}\quad
\left(\begin{array}{c} u\\ d\\ \end{array}\right)\quad{\rm quarks},
\\
N^{-2} &=& 2R^3j_0^2(x)[\Omega(\Omega-1) + Rm^*_q/2]/x^2,
\\
\beta &=& \sqrt{(\Omega-Rm^*_q)/(\Omega+Rm^*_q)},
\\
\Omega &=& \sqrt{x^2+(Rm^*_q)^2},
\qquad m^*_q = m_q - g^q_\sigma \sigma(\vr).
\ear

Once we have the quark wave functions in medium,
we can calculate the corresponding parton distributions.
All that are required are suitable modifications of the free space
expressions, which we now summarise.

In the case where the proton is described by
the MIT bag model with just three valence quarks in the 1s-state,
the dominant contribution to the twist-2 quark distribution is given by
\cite{tony91}:
\be
q_v(x) = \frac{M}{(2\pi)^3}\int d^3p_n \frac{|\phi_2 (\vec p_n)|^2}
{|\phi_3 (0)|^2} \delta(M(1-x) - p_n^+) |\tilde \psi_+ (\vec p_n)|^2.
\la{21}
\ee
This contribution comes from the case where the spectators to the hard
collision are two valence quarks, again in the 1s-state of the bag,
which can form a scalar (total spin S=0) or
a vector (S=1) system.
The $\phi$ factors come from the Peierls-Yoccoz
momentum projection which is used to build translationally invariant
states.
The integration is over the momentum of the diquark spectator
system in the intermediate state, $\psi_+$ is the plus component
of the quark wave function in momentum space, and in Eq.(\ref{21}) 
and in what follows $x$ is the standard 
Bjorken scaling variable.
More details concerning Eq.(\ref{21}) can be found in Ref.
\cite{tony91}.

After integration over the transverse components of
the momentum of the diquark pair, we have:
\be
q_v(x) = \frac{M}{(2\pi)^2}\int_{\frac{M^2(1-x)^2 - M_n^2}{2M
(1-x)}}^\infty
dp_n \frac{|\phi_2 (\vec p_n)|^2}{|\phi_3 (0)|^2}|\tilde
\psi_+ (\vec p_n)|^2.
\la{22}
\ee
In this paper, we will be working with the bag model wave function
for a massless quark, which is given by Eq.(\ref{2}) with
$V_\sigma = V_\omega = V_\rho = 0$
for the free space case. In its simplest form, the
bag calculation has just a few free parameters: the bag radius (for a
free nucleon),
the scalar ($M_s$) and vector ($M_v$) diquark masses, and the starting
scale, $\mu$, for the QCD evolution (at which
the bag model is supposed to best represent the non-perturbative
structure of the nucleon). They are fixed by
fitting the valence distribution to the respective existing
parametrizations
for the experimental data. This procedure has been quite successful in
the past \cite{tony91,me94a}, showing that the bag model
is able to describe not only the unpolarized valence sector, but also
has very good predictive power for the polarized sector \cite{me94b}.
In this work, we use $\mu^2 = 0.1\; GeV^2$, $R=0.8\; fm$, $M_s = 700\;
MeV$
and $M_v = 900\; MeV$, which
are the values for the parameters which give a good fit to the MRSA
parametrization of the valence distribution.

In calculating the twist-2 parton distributions in the bound nucleon one
must omit any interaction between the struck quark and the nuclear
medium.
However, the bound nucleon itself and the pair of spectator valence
quarks in the struck nucleon still feel the mean scalar and vector
potentials \cite{QMCSF}. That is, in-medium
the $\sigma$ and $\omega$ coupling at the quark level change the mass
and energy of the nucleon and the zero component of the
momentum of the intermediate state in the following way:
\bear
M &\rightarrow & M^* + 3V_\omega^q \nonumber \\*
p_n^0 &\rightarrow & (p_n^0)^* + 2V_\omega^q.
\la{23}
\ear
Note that the Bjorken variable, $x$, is defined in terms of the free
nucleon mass, $M$, so that $Mx$ is actually mass independent.
This is an important observation as it directly effects the
integration region.
\begin{figure}[htb]
\epsfig{file=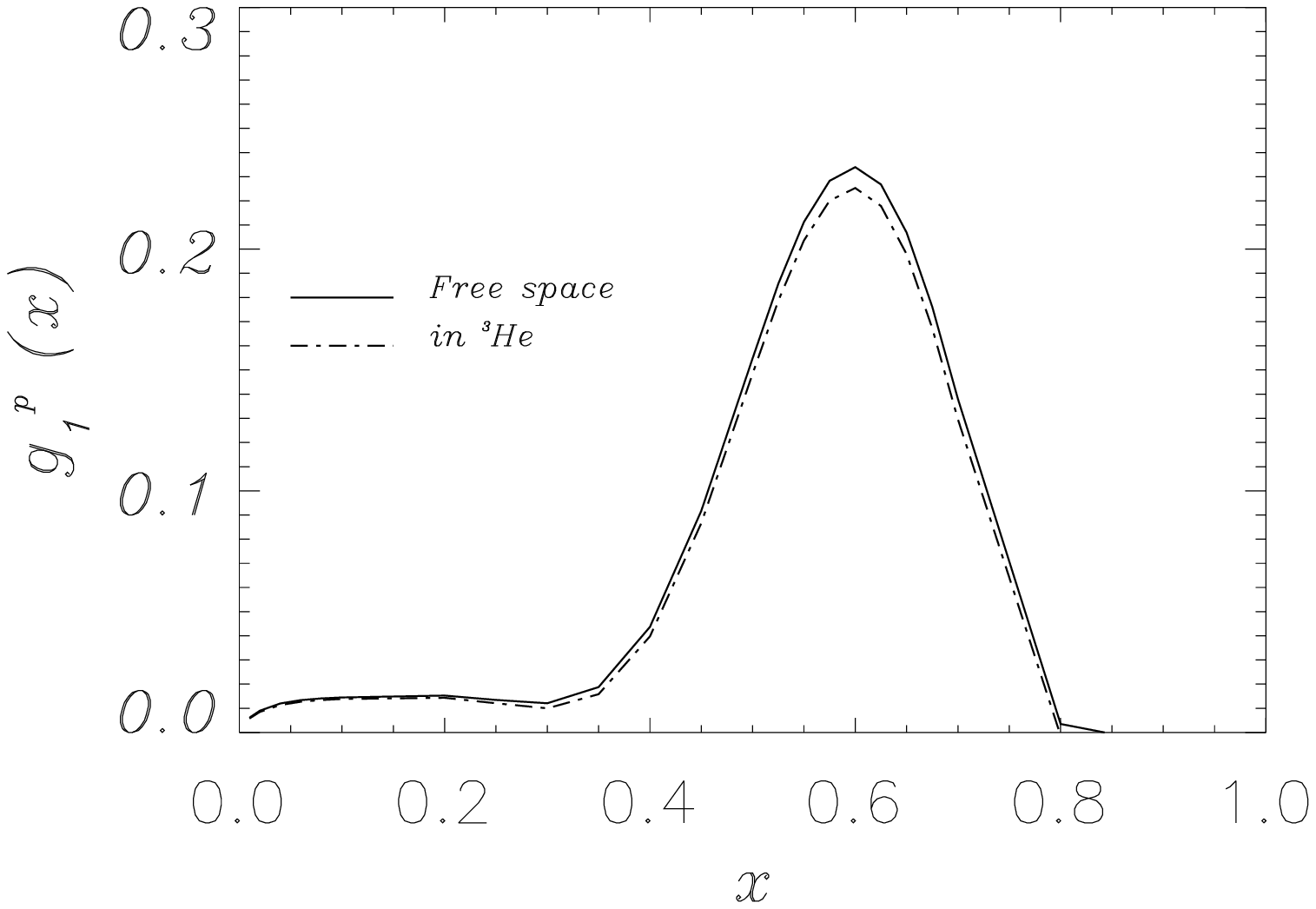,height=7cm}\,\quad
\epsfig{file=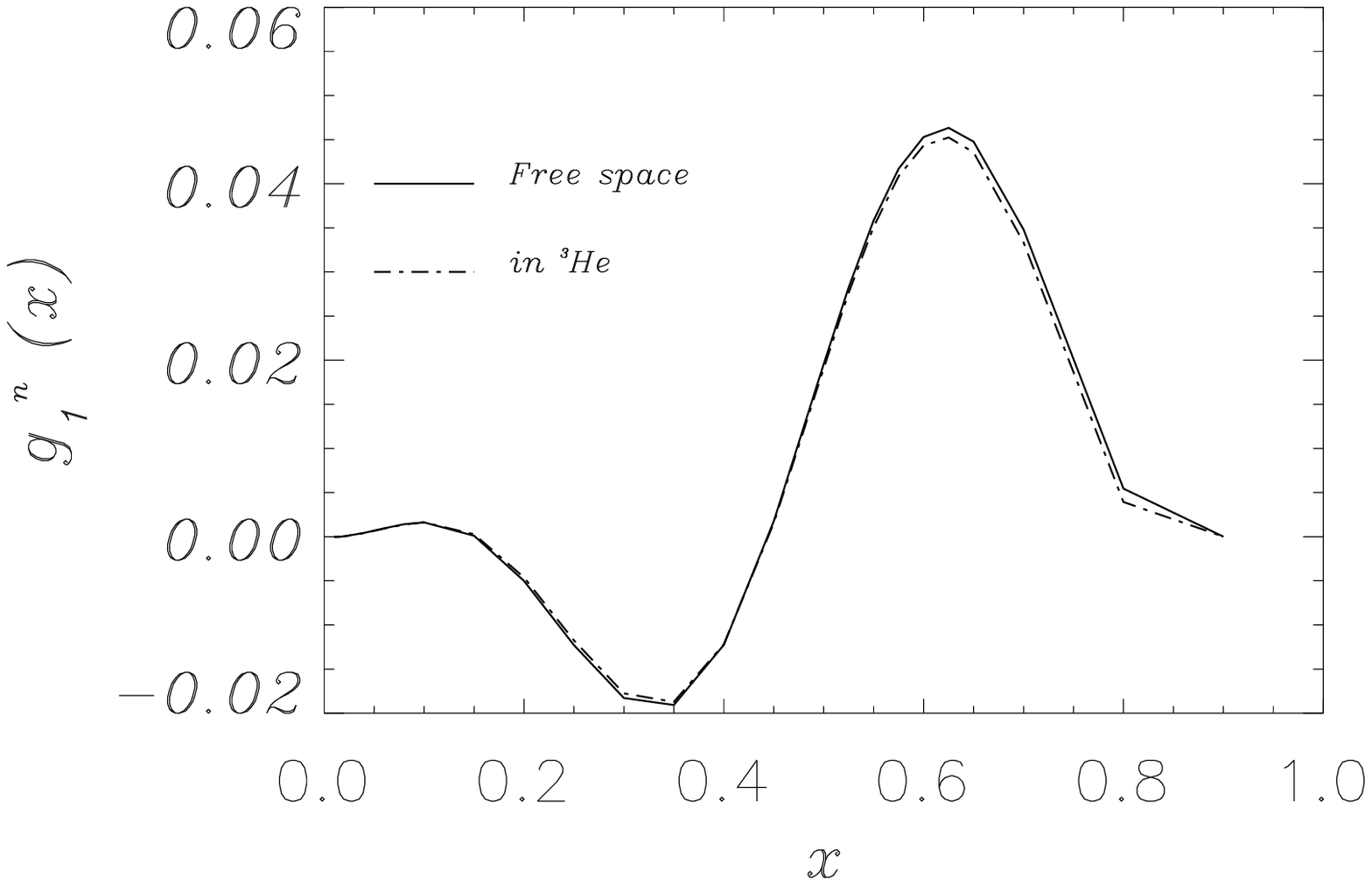,height=7cm}
\caption{The valence quark contributions to the
proton and neutron structure functions calculated in
free space using Eq. (\protect\ref{22}) and in $^3$He
using Eq. (\protect\ref{32}).}
\label{fig1}
\end{figure}

Incorporating these changes, the quark distribution for a nucleon at
rest in the medium is
given by:
\be
q_v^{(\tilde N)}(x) = \frac{M \tilde y}{(2\pi)^2}
\int_{\frac{(M^* + V_\omega^q - Mx)^2 - M_n^{* 2}}{2 (M^* + V_\omega^q -
Mx)}}^\infty
dp_n \frac{|\phi_2 (\vec p_n)|^2}{|\phi_3 (0)|^2}|\tilde
\psi_+ (\vec p_n)|^2,
\la{24}
\ee
with
\be
\tilde y \equiv \frac{M^* + 3V_\omega}{M} = \frac{p_{\tilde N}^0}{M}.
\la{26}
\ee
There is a second important observation to be made regarding the
in-medium
calculations. This concerns the $\rho$ meson coupling to the quarks
\cite{finite0,finite1,qhd}.
In this case, Eqs.(\ref{23}) are modified to
$M \rightarrow  M^* + 3V_\omega^q + \frac{1}{2}\tau_3^N V_\rho^N$ and
$p_n^0 \rightarrow  (p_n^0)^* + 2V_\omega^q + \sum_{j=1}^2\frac{1}{2}
\tau_3^q(j)V_\rho^q$,
where $\tau_3$ are the nucleon and quark isospin matrices and
$V_\rho^N = V_\rho^q$ is the rho meson potential.
We note that, with the inclusion of the $\rho$ meson,
these expressions imply different
integration minima for the $u$ and $d$ quark distributions in the bound
proton. Moreover, in a nucleus with $N \neq Z$, the valence
distributions
are no longer charge symmetric between the proton and the neutron,
meaning
that the substitutions $d_v^p (x) = u_v^n (x)$ and $u_v^p (x) = d_v^n
(x)$
may be less accurate in-medium than in free space.
This is a model independent phenomenon, and its origin,
unlike the free space investigations of isospin breaking in the
valence distributions \cite{londergan}, is not in the quark masses.
In practice, the mean field potential $V_\rho^q$ is quite a bit smaller
than those associated with $\sigma$ and $\omega$ and we shall drop it in
this first investigation.
\begin{figure}[htb]
\epsfig{file=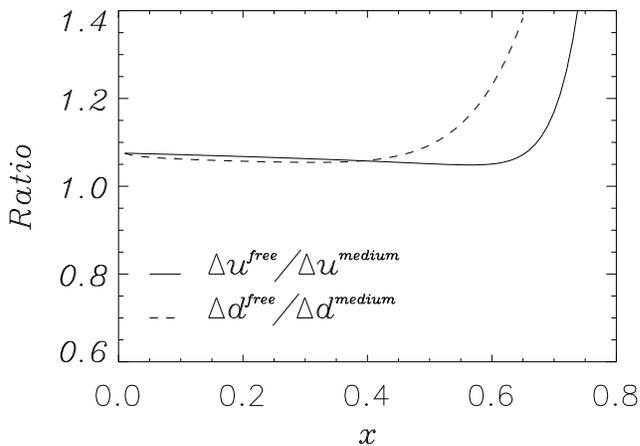,height=7cm}
\caption{The ratio of free to $^3$He polarized, valence quark
distributions in
the proton for the $u$ and $d$ quarks.}
\label{fig2}
\end{figure}

Although correct, Eq.(\ref{24}) is not yet suitable for the
in medium calculation for the following reason. When calculating
the quark distributions in a nucleon, we express them as a function
of $x$, which is a fraction of a chosen unit of momentum (in this
case the mass of the proton, $M$).
If we now want to calculate the quark distribution of
a nucleon which itself has a certain momentum distribution inside a
nucleus,
then this quark distribution will be a function of the momentum fraction
$y$ carried by the nucleon.
However, our Eq. (\ref{24}) was derived for a nucleon with momentum
$p_{\tilde N}^0 = M^* + 3V_\omega$.
We therefore define
a new valence quark momentum distribution:
\be
q_{v/\tilde N}(x) =  q_v^{(\tilde N)} (\tilde y x).
\la{29}
\ee
{}For an arbitrary nucleus, we
then have the following convolution formula for the nuclear valence
distribution at leading twist \cite{emcth,BT}:
\be
q^{(A)} (x) = \int_x^A \frac{dy}{y} f_{N/A}(y) q_{v/\tilde
N}(\frac{x}{y}).
\la{30}
\ee
where $f_{N/A}(y)$ is the usual momentum distribution function for
the nucleons inside the nucleus\cite{emcth,BT}.
This choice ensures the valence quark number (Gross Llewellyn Smith)
sum rule as well as
the appropriate momentum sum rule.
The distributions presented in
this paper will be for $q_{v/\tilde N} (x)$ calculated through
Eq.(\ref{29}). We emphasise that Eq.(\ref{30}) is the
standard nuclear convolution, which accounts for the kinematic
corrections associated with binding and Fermi motion,
so the ratio $q_{v/\tilde N}(x)/q_v(x)$ is a measure of the genuine
off-shell correction to the valence parton distribution.
Our parametrizations for this ratio (see Eqs. (\ref{35}) and (\ref{36})
below) can therefore be used in conjunction with any conventional
calculation of the nucleon momentum distribution, $f_{N/A}(y)$.

The quark distributions in medium are not only a function of
$x$ but also a function of the distance, $r$, from the centre of
the nucleus.
Hence we average the quark distributions
over the nuclear density:
\be
q_{v/\tilde N} (x) = \int d^3r q_{v/\tilde N} (x, r)
\rho_{\tilde N} (r).
\la{32}
\ee
Note that as we have already specified the momentum of the bound nucleon,
this averaging procedure amounts to a semi-classical approximation.
It is an interesting challenge for future work to improve on this point.
Here $\rho_{\tilde N} (r)$ is the probability density for the struck
nucleon and $q_{v/\tilde N} (x, r)$ is the quark distribution calculated
at the local nuclear density at $\vr$ for a bag with the effective mass,
radius and so on given by the QMC equations.
Eq. (\ref{32}) is valid for $^3$He, where the two protons and the
neutron
are in s-waves. However, $^6$Li may be thought of as an $\alpha$
particle plus a neutron-proton pair, each one in a p-state. It is this
pair,
with deuteron quantum numbers, which gives
the spin of the $^6$Li nucleus. Thus, when calculating
the $^6$Li spin structure function we average only over the
proton and neutron p-state densities  (calculated with
the help of the QMC equations). In the particular case of
the proton in $^6$Li, $\rho_{\tilde N}$ is simply the density of
protons in the p-state. For $g_1^n$ it is the neutron density in the
p-state.
In actual calculations, the structure function for the bound nucleon was
calculated over a grid of values of $r$, with interval 0.04 $fm$,
running
from $r = 0$ to $r = 4$ $fm$ for $^3$He, and from $r = 0$ to $r = 12 $
$fm$ for $^6$Li.
\begin{figure}[htb]
\epsfig{file=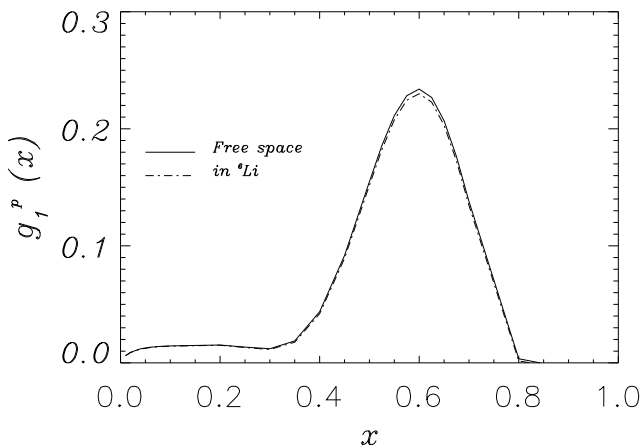,height=7cm}
\caption{The valence quark contribution to the
proton spin structure function calculated in
free space and in $^6$Li.}
\label{fig3}
\end{figure}

{}For the $^3$He case, we use \cite{3he} $\rho_{^3He} (r) =
0.149\; exp[-0.4244\; r^2]$ in Eq. (\ref{32}) to calculate the
corresponding distributions.
The results are shown in Fig.\ref{fig1}, where
we show the structure function, $g_1(x)$,
for a free proton (neutron) in comparison with that
for a proton (neutron) bound in $^3$He. Note that we have not carried
out the convolution (c.f. Eq.(\ref{30})) needed to include conventional
binding and Fermi motion effects, so the difference between the two
curves is solely a measure of the off-shell corrections for the bound
nucleon.
We see that these corrections are not significant in comparison with the
current experimental errors. This is
an important result when we remember that $^3$He has been
chosen because, from the point of view of its spin structure
function, it is assumed to be a pure neutron target.
Our result vindicates this view, since we get almost no correction
for the neutron spin structure function.
(On the other hand, our result says nothing new about the standard
nuclear
corrections associated with the kinematics of binding and Fermi motion,
nor about the nuclear structure corrections associated with wave
function components other than the neutron with a pair of spectator
protons in the $^1S_0$ state. Those corrections can be estimated from
standard nuclear structure calculations \cite{emcth,BT}.)
\begin{figure}[htb]
\epsfig{file=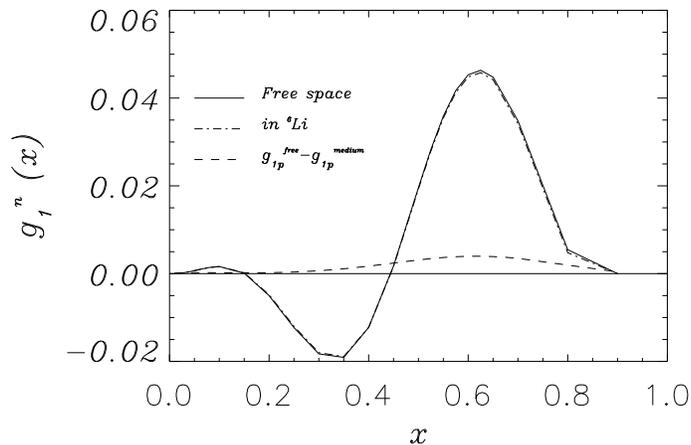,height=7cm}
\caption{The valence quark contribution to the
neutron spin structure function calculated in
free space and in $^6$Li. Also shown is the difference
$g_{1p}^{free}(x)$ - $g_{1p}^{^6Li}(x)$, which is relevant
for the extraction of $g_{1n}(x)$ from the measured value of
$g_1^{^6Li}$.}
\label{fig4}
\end{figure}

In Fig. \ref{fig2} we show the ratios of the free to the bound
polarized $u$ and $d$ quark, valence distributions. The in medium
corrections are practically the same for the $u$ and $d$ distributions
in the whole valence region, up to $x\approx 0.45$. In the
large $x$ region, the $d$ distribution in $^3$He is supressed
relative to its free counterpart at a faster rate than the $u$
distribution. The calculated ratio of the polarized, valence
quark distributions for the proton in free space to that in medium
can be parametrized in the following way:
\be
\frac{\Delta q_v (x)}{\Delta q_{v/\tilde N} (x)} =
a_q x^{b_q} + c_q x^{d_q} (1 - x)^{e_q},
\la{33}
\ee
with the parameters:
\bear
a_u = 118.41 &,&   a_d =  8.964    \nonumber \\*
b_u = 18.97  &,&   b_d =  7.5848   \nonumber \\*
c_u = 1.0758 &,&   c_d =  1.0515   \nonumber \\*
d_u = 0      &,&   d_d =  -0.0048  \nonumber \\*
e_u = 0.0335 &,&   e_d =  0.01086.
\la{35}
\ear

A similar procedure is used to calculated the in-medium corrections
to the proton and neutron spin structure function in a $^6$Li
target -- see Figs. \ref{fig3} and \ref{fig4}.
The main differences with respect to the $^3$He case are that the
nuclear density
in $^6$Li is calculated self-consistently in QMC, as
explained in the introduction, and the average is over the proton
and neutron p-states only.
The ratio of the free to bound,
polarized valence quark distributions can be parametrized as:
\be
\frac{\Delta q_v (x)}{\Delta q_{v/\tilde N} (x)} =
a_q + b_q x^{0.5} + c_q x + d_q x^{1.5} + e_q x^2 + f_q x^{2.5} + g_q
x^3,
\la{351}
\ee
with
\bear
a_u = 1.1267    &,&   a_d = 1.0904    \nonumber \\*
b_u = -2.2012   &,&   b_d = -1.4308   \nonumber \\*
c_u = 17.3645   &,&   c_d = 12.0596   \nonumber \\*
d_u = -66.6193  &,&   d_d = -50.0592  \nonumber \\*
e_u = 132.6824  &,&   e_d = 107.9883  \nonumber \\*
f_u = -131.2261 &,&   f_d = -116.1088 \nonumber \\*
g_u = 50.7843   &,&   g_d = 49.2399.
\la{36}
\ear
As in the case of $^3$He, the off-shell corrections to be made
to the deconvoluted nucleon spin structure functions are not large.
However, in order to extract $g_1^n (x, Q^2)$ from a $^6$Li
target, one has to subtract $g_1^p (x, Q^2)$ from the
measured value of $g_1^{^6Li}$. In this case, one can see from Fig.
\ref{fig4} that the use of the free $g_1^p (x, Q^2)$ would
induce some underestimate of $g_1^n (x, Q^2)$ over
the region $0.4 < x < 0.8$. The off-shell corrections
to the proton spin structure functions are as much as 6-8\% of $g_1^n$
itself. They will therefore become important to the determination of
$g_1^n (x, Q^2)$ when the experimental precision gets to this level.

Although the magnitude of the off-shell corrections calculated
here is model dependent, we believe that the
ratios should be less sensitive to the details of the
model. In this sense, the
parametrizations for $\Delta q_v(x)/\Delta q_{v/\tilde N}(x)$ through
Eqs. (\ref{33}) - (\ref{36}) are suitable for general use when
making the nuclear deconvolution. We note also that we
have verified that the ratios of the valence distributions presented
here are approximately scale independent so that they can be safely
used for $Q^2$ up to 10 GeV $^2$.

The results presented here are the first consistent calculations
of off-shell corrections to the proton and neutron valence spin
structure
functions in nuclei heavier than the deuteron.
By consistent we mean that the quark model used to
calculate the parton distributions is the same model used to calculate
the nuclear structure.
All the parameters were fixed in order to reproduce data other than
nuclear deep inelastic scattering and therefore the
results shown here really are predictions.
Most important, perhaps, is the fact that we have at our disposition
a procedure where the off-shell corrections to the
valence quark distributions can be systematically studied.

Overall, we have one main prediction together with one main
confirmation. First, it is reassuring that one can
indeed identify the measured spin structure function of
$^3$He with the neutron spin structure function, without
significant off-shell corrections. Second, our work indicates that
future experiments using $^6$Li as a target, will need to take
into account the in-medium corrections, of the type calculated here,
when extracting $g_1^n (x)$.
\newline

F.M.S. would like to thank the Special Research Centre for the
Subatomic Structure of Matter at the University of Adelaide for
support during a recent visit, where part of this work was performed.
This work was supported by the Australian Research Council and
by FAPESP (Brazil).

\addcontentsline{toc}{chapter}{\protect\numberline{}{References}}


\begin{thebibliography}{40}
\def\bi{\bibitem}
\bi{emcnuc} J. J. Aubert (EMC Collaboration), Phys. Lett. B 123
(1983) 275.
%
\bi{emcth} D. F. Geesaman, K. Saito and A. W. Thomas,
Annu. Rev. Nucl. Part. Sci.
45 (1995) 337; M. Arneodo, Phys. Rep. 240 (1994) 301.
%
\bi{EJ} J. Ellis and R. Jaffe, Phys. Rev. D 9 (1974) 1444;
Phys. Rev. D 10 (1974) 1669 (E).
%
\bi{emcspin} J. Ashman et al., Nucl. Phys. B 328 (1989) 1.
%
\bi{BJ} J. D. Bjorken, Phys. Rev. 148 (1966) 1467;
Phys. Rev. D 1 (1970) 1376.
%
\bi{smc} D. Adams et al., Phys. Rev. D 56 (1997) 5330.
%
\bi{e143} K. Abe et al., hep-ph/9802357.
%
\bi{D1} B. Adeva et al., Phys. lett. B 302 (1993) 533.
%
\bi{D11} D. Adams et al., Phys. Lett. B 357 (1995) 248;
Phys. Lett. B 396 (1997) 338.
%
\bi{D2} K. Abe et al., Phys. Lett. B 364 (1995) 61.
%
\bi{D3} K. Abe et al., Phys. Rev. Lett. 75 (1995) 25.
%
\bi{Deut} W. Melnitchouk, G. Piller and A. W. Thomas, Phys. Lett. B 346
(1995) 165; Phys. Rev. C 54 (1996) 894.
%
\bi{nform} H. Ankin {\em et al.}, Phys. Lett. B 336 (1994)
313; E. Bruins {\em et al.}, Phys. Rev. Lett. 75 (1995) 21;
M. Meyerhoff {\em et al.}, Phys. Lett. B 327 (1994) 201;
F. Klein, Proc. of PANIC96, Williamsburg (1996);
H. Schmieden, Proc. of SPIN96, Amsterdam (1996).
%
\bibitem{exnform} For an overview, see D. Drechsel {\em et al.},
Working Group Summary,  nucl-th/9712013, and references therein; and
also D.H. Lu, K. Tsushima, A.W. Thomas, A.G. Williams and K.
Saito, ADP-98-7/T286, nucl-th/9804009.
%
\bi{strfn} A. Le Yaouanc {\it et al.}, Phys. Rev. D 11 (1975) 2636.
%
\bi{RJ}
R. L. Jaffe, D 11 (1975) 1593.
%
\bi{PP} R. Parisi and G. Petronzio,
Phys. Lett. B 93 (1976) 331.
%
\bi{GRV}
M. Gl\"uck, E. Reya and A. Vogt,
Z. Phys. C 67 (1995) 433; hep-ph/9806404.
%
\bi{ST1} A. I. Signal and A. W. Thomas, Phys. Lett. B 211 (1988)
481; Phys. Rev. D 40 (1989) 2832.
%
\bi{tony91} A. W. Schreiber, A. I. Signal and A. W. Thomas,
Phys. Rev. D 44 (1991) 2653; A. W. Schreiber {\it et al.},
Phys. Rev. D 45 (1992) 3069.
%
\bi{me94a} F. M. Steffens and A. W. Thomas, Prog. of Theor. Phys. Suppl.
120 (1994) 145.
%
\bibitem{gui}P.A.M. Guichon, Phys. Lett. B 200 (1988) 235.
%
\bibitem{finite0}P.A.M. Guichon, K. Saito, E. Rodionov and A.W. Thomas,
Nucl. Phys. A 601 (1996) 349;\\
P.A.M. Guichon, K. Saito and A.W. Thomas, nucl-th/9602022,
Australian Journal of Physics 50 (1997) 115.
%
\bibitem{finite1}K. Saito, K. Tsushima and A.W. Thomas,
Nucl. Phys. A 609 (1996) 339.
%
\bibitem{qmcapp}K. Saito, K. Tsushima and A.W. Thomas,
Phys. Rev. C 55 (1997) 2637; {\it ibid} C 56 (1997) 566;
Phys. Lett. B 406 (1997) 287; Mod. Phys. Lett. A 13 (1998) 769.
%
\bibitem{blu}P.G. Blunden and G.A. Miller, Phys. Rev. C 54 (1996) 359.
%
\bi{Jen}
X. Jin and B.K. Jennings, Phys. Lett. B 374 (1996) 13;
Phys. Rev. C 54 (1996) 1427; {\it ibid} C 55 (1997) 1567;\\
H. M\"{u}ller and B.K. Jennings, Nucl. Phys. A 626 (1997) 966;\\
H. M\"{u}ller, Phys. Rev. C 57 (1998) 1974.
%
\bibitem{qhd}J.D. Walecka, Ann. Phys. (N.Y.) 83 (1974) 491; \\
B.D. Serot and J.D. Walecka, Adv. Nucl. Phys. 16 (1986) 1.
%
\bi{3he} Roger C. Barret and Daphne F. Jackson, in ``Nuclear Sizes
and Structure'', Clarendon Press, Oxford 1977; K. Saito, K. Tsushima
and A. W. Thomas, Phys. Rev. C 56 (1997) 566.
%
\bi{me94b} F. M. Steffens, H. Holtmann and A. W. Thomas, Phys. Lett. B
358 (1995) 139.
%
\bi{QMCSF} A. W. Thomas {\it et al.}, Phys. Lett. B 233 (1989) 43;
K. Saito and A. W. Thomas, Nucl. Phys. A 574 (1994) 659.
%
\bi{londergan} J. T. Londergan and A. W. Thomas, Progress in Particle
and Nuclear Physics, 41 (1998) 49.
%
\bi{BT} R. P. Bickerstaff and A. W. Thomas, J. Phys. G 15 (1989) 1523.
%

\end{thebibliography}
\end{document}